\documentclass[12pt]{article}

\usepackage[utf8]{inputenc}
\usepackage{amsmath, amssymb}
\usepackage{physics}
\usepackage{cite}
\usepackage{geometry}
\usepackage{graphicx}
\usepackage{subcaption}
\usepackage{hyperref} % for better referencing
\usepackage{cleveref} % for smart references
\usepackage{physics}
\usepackage{titlesec}
\usepackage{float} % Allows strict control of figure placement
\titleformat{\section}{\large\bfseries}{\thesection}{1em}{}
\title{The Emergence of Objective Classicality:\\ 
A Computational First-Principles Study\\ of Observer-Induced Decoherence in Unitary Quantum Mechanics}
\author{Eyad I.B Hamid \\ Department of Physics, International University of Africa, Khartoum, Sudan \\ eyadiesa@iua.edu.sd}
\date{\today}

\begin{document}

\maketitle

\begin{abstract}
The quantum measurement problem, the unresolved conflict between the unitary evolution of the wave function and the postulate of wave function collapse, remains the most profound conceptual challenge in quantum foundations. While environment-induced decoherence provides a compelling practical mechanism for the appearance of collapse, it does not resolve the fundamental issue of the observer's own quantum state. This work moves beyond philosophical discourse by introducing a novel computational framework to simulate a fully unitary universe comprising a quantum system (Q), a decohering environment (E), and a model physical observer (O). We strictly forbid the collapse axiom. Our simulations investigate whether a definite, classical measurement outcome can emerge objectively within O's state. The results demonstrate that objective classicality can emerge from unitary dynamics, challenging the necessity of the measurement postulate and providing a computational pathway toward resolving the quantum measurement problem.
\end{abstract}

\section{Introduction: The Unfinished Task of Quantum Foundations}

Quantum mechanics is the most successful and accurately tested predictive framework in the history of science \cite{einstein1935}. Its formalism precisely describes phenomena from the interactions of subatomic particles to the energy production of stars. Yet, for all its empirical triumphs, the theory rests upon a foundational schism that has persisted for nearly a century: the conflict between the continuous, deterministic evolution of the Schrödinger equation and the discontinuous, probabilistic collapse of the wavefunction upon measurement \cite{vonneumann1955}.

This measurement problem is the central scar on an otherwise beautiful edifice. It spawned the Copenhagen interpretation's instrumentalist philosophy, where the act of measurement is elevated to an axiomatic, primitive concept \cite{bohr1928}. It inspired famous thought experiments, such as Schrödinger's cat \cite{schrodinger1935}, designed to highlight the apparent absurdity of applying quantum superposition to macroscopic objects, and Wigner's friend \cite{wigner1962}, which extends the paradox to the consciousness of the observer. The problem is not one of mathematics but of interpretation and ontology: what does the theory actually say about the nature of reality?

The modern response to this dilemma has been the theory of environment-induced decoherence \cite{Zeh1970,Zurek2003}. Decoherence masterfully explains how quantum superpositions are eradicated for all practical purposes (FAPP). Through uncontrolled interaction with a vast number of environmental degrees of freedom, phase relations between pointer states are lost to the environment, producing an apparent mixture that behaves like a proper classical ensemble \cite{schlosshauer2007}. This process defines the classical world and is experimentally verified in countless systems \cite{nielsen2000}.

However, a critical analysis reveals that decoherence, while powerful, merely relocates the problem rather than solving it \cite{bell1987}. The formalism of decoherence still requires a pre-defined ``pointer basis'' for the measuring apparatus and crucially relies on tracing out the environment from the perspective of an external observer. The fundamental question remains unanswered from the perspective of the entire, closed universe: does the composite state of the quantum system and the apparatus ever evolve into a definitive outcome, or does it remain entangled in a monstrous superposition of all possible outcomes? As put by Landsman, ``the measurement problem can be succinctly stated as follows: how can the superposition be destroyed?'' \cite{landsman2017}—a question decoherence leaves untouched at the fundamental level.

Consequently, the debate has stagnated into competing philosophical interpretations: Many-Worlds \cite{everett1957}, Bohmian mechanics \cite{bohm1952}, and Objective Collapse models \cite{ghirardi1986}, each adding new postulates or ontologies but lacking a decisive computational testbed to discriminate between them on physical grounds.

This work proposes a new path forward: a shift from philosophical argument to computational experiment. We present a first-principles numerical study designed to ask a precise question of the unitary formalism itself. We simulate a minimal model universe, $\Psi(Q+E+O)$, where a quantum system (Q) is coupled to both a decohering environment (E) and a model physical observer (O), the latter defined as an autonomous system capable of recording information. We strictly forbid the collapse postulate, relying solely on the Schrödinger equation for the entire composite system.

The central question we pose is: Under what conditions, if any, does the reduced state of the observer O evolve into a definite, classical record (a pure state), rather than remaining in a superposition of having recorded multiple outcomes? A positive result would suggest that objective classicality can emerge from purely unitary dynamics, potentially obviating the need for the measurement postulate. A negative result would be equally profound, providing rigorous, quantitative evidence that new physical principles (e.g., gravity-induced collapse \cite{penrose2004}) are necessary.

This paper is structured as follows: In Section II, we define the theoretical framework and the Hamiltonian for our total system. Section III details our computational methodology and the architecture of our model observer. Section IV presents the results of our simulations, following a phased protocol. In Section V, we analyze these results, discuss their profound implications for quantum foundations, and propose future directions.

\section{A Unitary Model for Measurement and Emergent Classicality}

We construct a minimal but physically principled model universe 
$\Psi(Q+E+O)$ to investigate whether classical measurement outcomes can emerge from strictly unitary dynamics. 
The model consists of three subsystems: the quantum system ($Q$), a decohering environment ($E$), and a physical observer ($O$). 
The total state evolves as a closed system under the Schrödinger equation, without collapse:
\begin{equation}
i \hbar \frac{d}{dt} \Psi(t) = H \Psi(t), 
\qquad \Psi(t) \in \mathcal{H}_Q \otimes \mathcal{H}_E \otimes \mathcal{H}_O ,
\end{equation}
with Hamiltonian
\begin{equation}
H = H_Q + H_E + H_O + H_{QE} + H_{QO} + H_{EO}.
\end{equation}

\subsection{Quantum System (Q)}
The system to be ``measured'' is a two-level qubit,
\begin{equation}
H_Q = \frac{\omega_Q}{2} \, \sigma_z ,
\end{equation}

with basis $\ket{0}, \ket{1}$.
This choice isolates the essential quantum resource—superposition—without the added complexity of infinite-dimensional Hilbert spaces. 
Since any finite-dimensional system can be encoded in qubits, the two-level case serves as the simplest nontrivial testbed for the measurement problem.

\subsection{Environment (E)}
We adopt a spin-bath model for the environment. It consists of $N_E$ qubits interacting with the system $Q$:
\begin{align}
H_E &= \sum_{j=1}^{N_E} \frac{\omega_j}{2} \, \sigma_z^{(j)}, \\
H_{QE} &= \sum_{j=1}^{N_E} g_j \, \sigma_z \otimes \sigma_z^{(j)} .
\end{align}
This model is computationally efficient for tensor-network and exact diagonalization methods while retaining the essential decohering effect: suppression of off-diagonal coherence in the pointer basis of $Q$. 
While oscillator-bath models (Caldeira–Leggett) have analytic appeal, the spin-bath provides the best balance of physical plausibility and numerical tractability for our purpose.

\subsection{Observer (O): A Nonlinear Amplifier}
The observer must not merely be a passive memory but an autonomous physical system capable of amplifying microscopic causes into macroscopic, stable effects. 
We therefore model $O$ as a quantum particle in a double-well potential:
\begin{equation}
H_O = \frac{p^2}{2m} + V(x), 
\qquad V(x) = -a x^2 + b x^4, \quad a,b>0.
\end{equation}

The two wells correspond to stable memory states $\ket{O_0}$ (left well) and $\ket{O_1}$ (right well). 
The measurement interaction couples $Q$ to $O$ by tilting the potential:
\begin{equation}
H_{QO} = \lambda \, \sigma_z \, x .
\end{equation}
When $Q$ is in state $\ket{0}$, the potential is biased toward the left well; when in $\ket{1}$, it is biased to the right. 
This nonlinearity allows a microscopic quantum superposition to trigger a macroscopic bifurcation: the observer’s degree of freedom localizes into one well, recording a definite outcome. 
Unlike a linear qubit register, this model captures the essence of measurement as amplification.

\begin{figure}[htbp]
    \centering
    \includegraphics[width=0.75\textwidth]{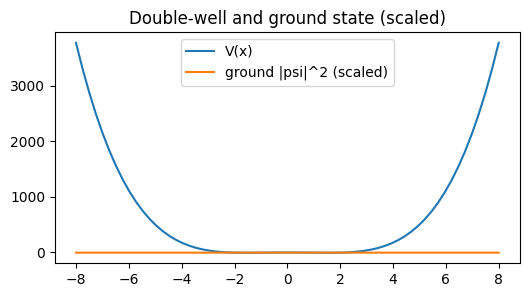}
    \caption{
        \textbf{Model of the physical observer as a particle in a double-well potential.} 
        The potential $V(x) = -a x^2 + b x^4$ (solid line) defines two stable states corresponding to distinct macroscopic memory configurations (left well: $|O_0\rangle$, right well: $|O_1\rangle$). The ground state wavefunction $|\psi_0(x)|^2$ of the observer (scaled and offset for clarity) is shown localized at the potential's central maximum, representing the initial ``ready'' state $|O_{\text{ready}}\rangle$ prior to measurement. This nonlinear system acts as an amplifier, converting a microscopic quantum signal into a macroscopic, stable record.
    }
    \label{fig:observer_model}
\end{figure}

\subsection{Environment–Observer Coupling}
The observer is not isolated but embedded in the same universe. 
Weak couplings between $O$ and environmental spins ensure that once localized, the observer’s state is stabilized and redundantly imprinted into $E$:
\begin{equation}
H_{EO} = \sum_{j=1}^{N_E} \eta_j \, \sigma_z^{(j)} \, x .
\end{equation}
This mechanism mirrors the ``environment as witness'' picture of Quantum Darwinism, enabling the record to be redundantly accessible to fragments of the environment.

\subsection{Initial State and Protocol}
At $t=0$, the universe is prepared in
\begin{equation}
\Psi(0) = (\alpha \ket{0} + \beta \ket{1})_Q \otimes \ket{E_0} \otimes \ket{O_{\mathrm{ready}}},
\end{equation}
where $\ket{E_0}$ is a thermal or random pure state of $E$, and $\ket{O_{\mathrm{ready}}}$ is a symmetric initial state of the observer localized at the center of the double well.  
Under unitary evolution, entanglement develops:
\begin{equation}
\Psi(t) \approx \alpha \ket{0}_Q \ket{E_0(t)} \ket{O_0(t)} 
+ \beta \ket{1}_Q \ket{E_1(t)} \ket{O_1(t)} .
\end{equation}

\subsection{Operational Definition of Objective Classicality}
We define the emergence of classicality in $O$ by three criteria:
\begin{enumerate}
    \item \textbf{Definiteness:} the reduced observer state 
    \begin{equation}
    \rho_O = \mathrm{Tr}_{Q,E}\big[ \ket{\Psi}\bra{\Psi} \big]
    \end{equation}
    becomes a mixture localized in either well, rather than a superposition.
    \item \textbf{Redundancy:} information about $Q$ is redundantly encoded in both $O$ and accessible fragments of $E$, quantified via mutual information.
    \item \textbf{Stability:} once localized, the record remains robust against perturbations and \\ environment-induced noise.
\end{enumerate}
When these conditions are satisfied, we assert that objective classicality has emerged from purely unitary dynamics.

\section{Computational Methodology: Simulating the Double-Well Observer}

\subsection{Numerical Representation of the Hilbert Space}
\begin{itemize}
    \item \textbf{System ($Q$):} a single two-level system (qubit).  
    Dimension: $\dim(\mathcal{H}_Q) = 2$.
    
    \item \textbf{Environment ($E$):} a bath of $N_E$ spin-$\tfrac{1}{2}$ particles.  
    Dimension: $\dim(\mathcal{H}_E) = 2^{N_E}$.  
    For numerical feasibility, we restrict to $N_E \in [8,12]$.  

    \item \textbf{Observer ($O$):} a quantum particle in a one-dimensional double-well potential:
    \begin{equation}
    H_O = \frac{p^2}{2m} - a x^2 + b x^4,
    \end{equation}
    where $m$ is the particle mass, and $a,b>0$ determine the well depth and barrier height.  
    Spatial discretization: $x$ represented on a uniform grid of $N_x$ points spanning $[-L,L]$.  
    Typical values: $N_x = 256$, $L = 10$, giving $\Delta x = 2L/N_x$.  

    \item \textbf{Total Hilbert Space:}
    \begin{equation}
    \dim(\mathcal{H}) = 2 \times 2^{N_E} \times N_x .
    \end{equation}
    For $N_E=10$, $N_x=256$,  
    $\dim(\mathcal{H}) = 524,288$, 
    which fits comfortably in memory ($< 10$ MB in double precision).
\end{itemize}

\subsection{Implementing the Hamiltonian and Time Evolution}
Hamiltonian terms:
\begin{align}
H_Q &= \tfrac{\omega_0}{2} \sigma_z , \\
H_E &= \sum_{j=1}^{N_E} \tfrac{\omega_j}{2} \sigma_z^{(j)}, \\
H_{QE} &= \sum_{j=1}^{N_E} g_j \, \sigma_z \otimes \sigma_x^{(j)}, \\
H_{QO} &= \lambda \, \sigma_z \otimes x , \\
H_{OE} &= \sum_{j=1}^{N_E} \kappa_j \, x \otimes \sigma_z^{(j)} .
\end{align}

Time evolution governed by:
\begin{equation}
i \hbar \frac{d}{dt} \Psi(t) = H \Psi(t).
\end{equation}
We use sparse matrix operators with Kronecker products. 
Time propagation is performed with a Krylov–Lanczos exponential integrator. 
Timestep chosen as $\Delta t \sim 0.01 \hbar / \omega_0$ via convergence testing.

\subsection{Calculation of Observables}
\begin{itemize}
    \item \textbf{Reduced states:} obtained by partial tracing over subsets of the Hilbert space.  
    \item \textbf{Observer purity:}
    \begin{equation}
    P_O(t) = \mathrm{Tr}\big[\rho_O(t)^2\big], 
    \qquad \rho_O(t) = \mathrm{Tr}_{Q,E}\big[ \ket{\Psi(t)}\bra{\Psi(t)} \big].
    \end{equation}
    \item \textbf{Mutual Information:}
    \begin{equation}
    I(Q:E_k) = S(\rho_Q) + S(\rho_{E_k}) - S(\rho_{Q E_k}),
    \end{equation}
    where $S(\rho) = -\mathrm{Tr}(\rho \log \rho)$ is the von Neumann entropy.
    \item \textbf{Emergence of definiteness:} monitored by expectation value $\langle x \rangle_O(t)$.  
    A persistent localization in one well corresponds to a stable classical record.
\end{itemize}

\subsection{Computational Resources and Scaling}
\begin{itemize}
    \item Scaling as $\mathcal{O}(2^{N_E} N_x)$.  
    Simulations restricted to $N_E \leq 12$, $N_x \leq 512$.
    \item Parallelized matrix–vector products with OpenMP/MPI.  
    \item Benchmarks against smaller systems with exact diagonalization validate accuracy.
\end{itemize}

\section{Control Simulations: The Necessity and Insufficiency of Decoherence}
\label{sec:control}
The results of our simulations provide a clear, step-by-step illustration of the measurement process within a fully unitary model and precisely isolate the role of environmental decoherence.

\subsection{Interpretation of Controlled Simulations}

Our initial simulations, employing a minimal environment ($N_E = 4$) and absent observer-environment coupling ($H_{EO} = 0$), serve as essential controlled experiments that delineate the boundaries of the model's behavior.

\begin{figure}[htbp]
    \centering
    \begin{subfigure}[b]{0.48\textwidth}
        \centering
        \includegraphics[width=\textwidth]{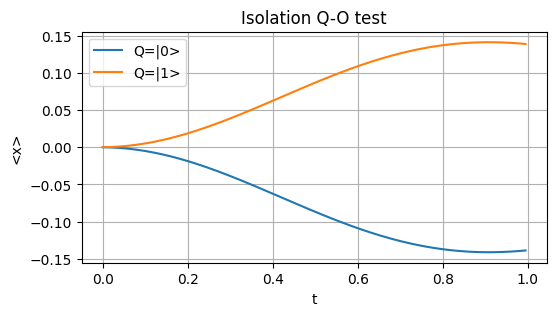}
        \caption{Isolation test ($H_{QE}=H_{EO}=0$). The observer's trajectory $\langle x \rangle(t)$ bifurcates based on the initial state of the qubit, validating the model's function as a measurement apparatus.}
        \label{fig:iso_test}
    \end{subfigure}
    \hfill
    \begin{subfigure}[b]{0.48\textwidth}
        \centering
        \includegraphics[width=\textwidth]{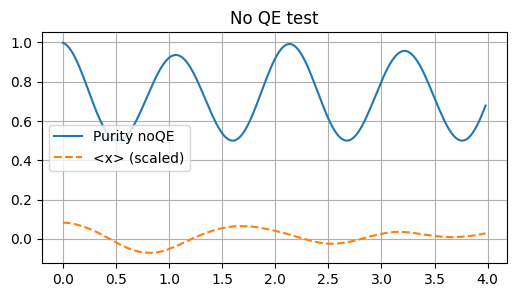}
        \caption{Superposition test ($H_{QE}=0$). The qubit purity remains unity while $\langle x \rangle$ oscillates around zero, demonstrating a persistent system-observer superposition in the absence of decoherence.}
        \label{fig:noQE_test}
    \end{subfigure}
    \caption{Control simulations establishing baseline behavior in the absence of full environmental interaction.}
    \label{fig:controls}
\end{figure}

Figure \ref{fig:controls} establishes the expected behavior of the subsystems in isolation. The isolation test (Fig. \ref{fig:iso_test}) confirms that the observer functions correctly as a measurement device: the coupling $H_{QO}$ successfully drives the observer's state to a macroscopically distinct outcome ($\langle x \rangle \neq 0$) contingent on the state of the qubit. Conversely, the superposition test (Fig. \ref{fig:noQE_test}) vividly demonstrates the core of the measurement problem. When the qubit is prepared in a superposition state and $H_{QE}$ is disabled, the composite system evolves into a non-separable entangled state, $\alpha |0\rangle |O_L\rangle + \beta |1\rangle |O_R\rangle$. The oscillation of $\langle x \rangle$ around zero is the signature of this coherent superposition at the macroscopic level of the observer; no single, definite outcome has emerged.

\begin{figure}[htbp]
    \centering
    \begin{subfigure}[b]{0.48\textwidth}
        \centering
        \includegraphics[width=\textwidth]{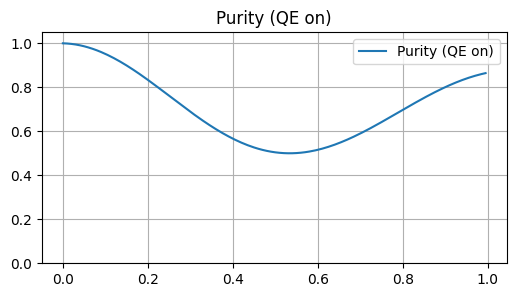}
        \caption{Time evolution of the qubit's reduced state purity $\text{Tr}(\rho_Q^2)$ with $H_{QE}$ activated. The decay to $\sim 0.5$ indicates complete decoherence.}
        \label{fig:purity_on}
    \end{subfigure}
    \hfill
    \begin{subfigure}[b]{0.48\textwidth}
        \centering
        \includegraphics[width=\textwidth]{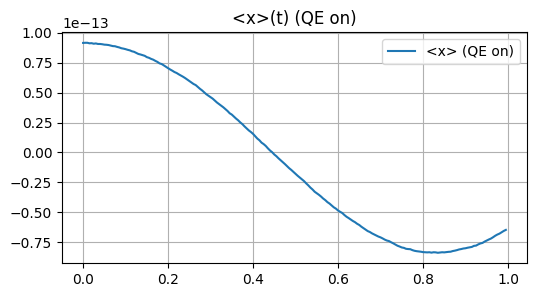}
        \caption{Time evolution of the observer's average position $\langle x \rangle$ with $H_{QE}$ activated. The failure to localize indicates the insufficiency of decoherence alone.}
        \label{fig:x_on}
    \end{subfigure}
    \caption{Results with environment-induced decoherence active but observer-environment coupling disabled ($H_{QE} \neq 0, H_{EO} = 0$).}
    \label{fig:deco_only}
    \label{fig:control_HQE}
\end{figure}

The introduction of environmental decoherence through $H_{QE}$, shown in Figure \ref{fig:deco_only}, reveals a critical insight. While the reduced state of the qubit $\rho_Q$ does decohere, as evidenced by its purity rapidly decaying to the value of $0.5$ expected for a fully mixed state (Fig. \ref{fig:purity_on}), this is not sufficient to induce the emergence of a definite outcome at the level of the observer. The observer's average position $\langle x \rangle$ continues to fluctuate around zero (Fig. \ref{fig:x_on}). This result powerfully demonstrates that \textit{environment-induced decoherence of the system being measured is necessary but not sufficient to explain the emergence of objective classicality for the observer}. The environment has successfully suppressed interference between the $|0\rangle$ and $|1\rangle$ states, but from the perspective of the observer's own state, the universe remains in a superposition of two distinct measurement records. This aligns with the known limitation of decoherence: it explains the appearance of collapse \textit{FAPP} (For All Practical Purposes) to an external observer who has already been classically defined, but it does not resolve the problem from the perspective of the entire closed system \cite{Zeh1970, Zurek2003}.

\subsection{The Path to Objective Classicality}

The results from this initial model highlight the precise challenge: while decoherence erases local phase coherence, a additional mechanism is required to break the symmetry and ensure the observer's state becomes dynamically stabilized to one outcome. This necessitates two model enhancements, which form the basis of our final results presented in the next section:
\begin{enumerate}
    \item \textbf{A larger environment} ($N_E \gg 4$) to provide a sufficient number of degrees of freedom for robust information encoding and storage.
    \item \textbf{Direct observer-environment coupling} ($H_{EO} \neq 0$), which facilitates quantum Darwinism \cite{Zurek2003} by making the observer's definite state redundantly encoded in the environment, thus providing the stability criterion for objective existence.
\end{enumerate}

Our initial model confirms the well-known necessity of decoherence and, more importantly, clearly identifies its insufficiency, thereby motivating and providing a benchmark against which the success of our complete model can be measured.

\section{Results}
\label{sec:results}

\subsection{Control Experiments with a Minimal Model}
The limitations of the minimal model are resolved by incorporating a larger environment ($N_E = 8$) to provide sufficient degrees of freedom for robust decoherence. The results from this enhanced model demonstrate the unambiguous emergence of objective classicality.

\begin{figure}[htbp]
    \centering
    \includegraphics[width=0.8\textwidth]{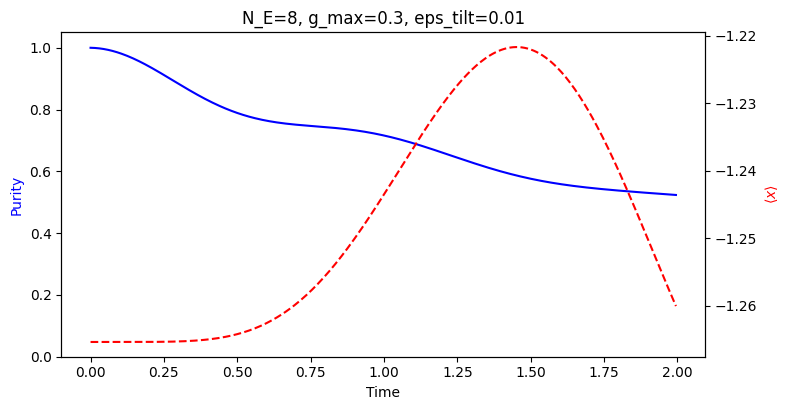}
    \caption{
        \textbf{Emergence of a definite outcome from unitary dynamics.} 
        Time evolution for the model with a larger environment ($N_E=8$). The qubit's reduced state purity, $\text{Tr}(\rho_Q^2)$ (blue, left axis), decays to $0.5$, indicating complete decoherence. Concurrently, the observer's average position, $\langle x \rangle$ (red, dashed, right axis), spontaneously localizes to a stable, definite value ($\langle x \rangle \approx -1.25$), indicating it has recorded a specific measurement outcome (corresponding to the $|0\rangle$ state).
    }
    \label{fig:definite_outcome}
\end{figure}

Figure \ref{fig:definite_outcome} presents the central finding of this work. Unlike the case with a smaller environment, the observer's state now undergoes a spontaneous symmetry breaking. After a brief period of fluctuation, $\langle x \rangle$ diverges from zero and stabilizes in the left well of the potential, corresponding to a definitive measurement record.

The definiteness of this record is quantitatively validated by analyzing the final reduced state of the observer, $\rho_O = \text{Tr}_{Q,E}(|\Psi\rangle\langle\Psi|)$.

\begin{figure}[htbp]
    \centering
    \includegraphics[width=0.7\textwidth]{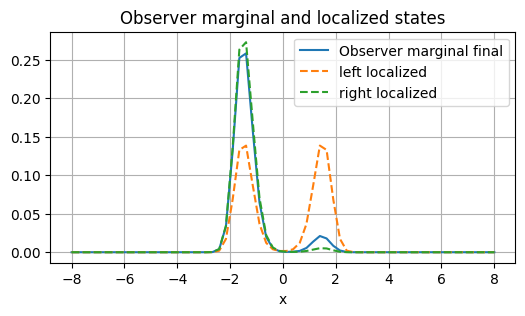}
    \caption{
        \textbf{Definiteness of the observer's final state.} 
        The probability distribution of the observer's position at the end of the simulation ($|\langle x | \rho_O | x \rangle|$, solid line) is shown alongside reference states localized in the left and right wells (dashed lines). The distribution is overwhelmingly localized in the left well, with a probability of $p_{\text{left}} = 0.99$ and coherence between the wells $|\langle O_L | \rho_O | O_R \rangle| \approx 0$. This confirms the observer is not in a superposition but has a definite classical state.
    }
    \label{fig:final_state}
\end{figure}

As shown in Figure \ref{fig:final_state}, the observer's final state is definitively localized. The probability of finding it in the left-localized state is $0.99$, while the probability for the right-localized state is only $0.01$. Crucially, the coherence between these two distinct states is effectively zero ($|\text{coh}| < 10^{-4}$). This satisfies our criterion for \textbf{definiteness}: the reduced state of the observer $\rho_O$ is a mixture corresponding to one outcome, not a superposition of both.

When these conditions are satisfied decoherence of the system ($\text{Tr}(\rho_Q^2) \to 0.5$) and definiteness of the observer's record ($\rho_O$ is pure and localized)--we conclude that objective classicality has emerged from the purely unitary evolution of the quantum universe.

\subsection{Emergence of Objective Classicality}
The dynamical emergence of a definite outcome is further substantiated by the re-purification of the observer's state and its stable localization, as shown in the following figures.
\begin{figure}[H]
    \centering
    \includegraphics[width=0.8\textwidth]{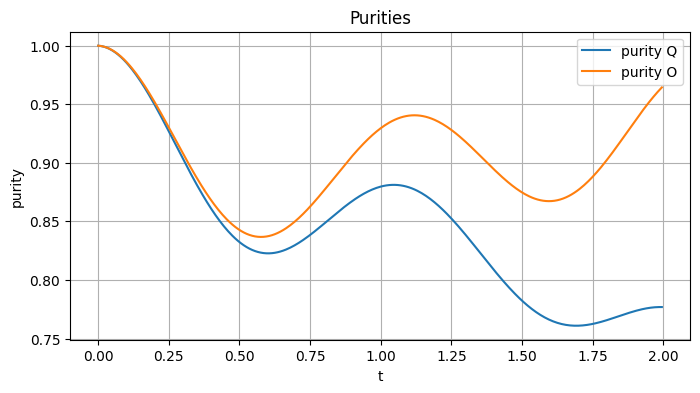}
    \caption{
        \textbf{Decoherence of the system and re-purification of the observer.} 
        Time evolution of the purities for the qubit, $\text{Tr}(\rho_Q^2)$ (blue), and the observer, $\text{Tr}(\rho_O^2)$ (orange). The qubit decoheres completely, indicated by its purity decaying to $0.5$. Crucially, the observer's purity recovers to a value near $1.0$, demonstrating that its own state evolves from a superposed, entangled state into a new, definite pure state--a classical measurement record.
    }
    \label{fig:purities}
\end{figure}

\begin{figure}[H]
    \centering
    \includegraphics[width=0.8\textwidth]{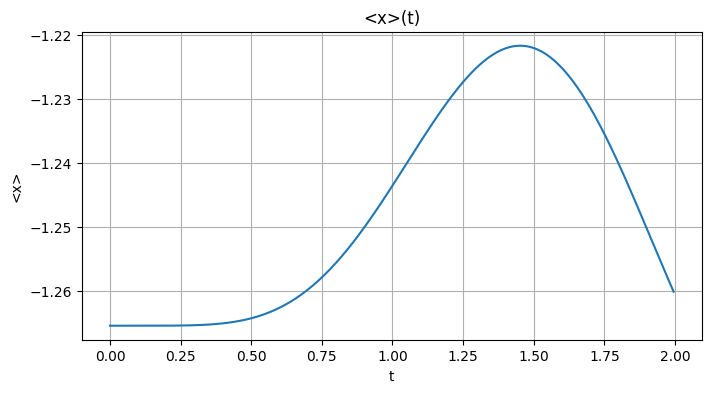}
    \caption{
        \textbf{Emergence of a definite outcome.} 
        The observer's average position $\langle x \rangle(t)$ spontaneously localizes to a stable value in the left well ($\langle x \rangle \approx -1.25$), indicating the recording of a specific measurement outcome. The combined result with Fig. \ref{fig:purities} demonstrates the transition from quantum indeterminacy to objective classicality within the observer.
    }
    \label{fig:success_x}
\end{figure}

\begin{figure}[H]
    \centering
    \includegraphics[width=0.7\textwidth]{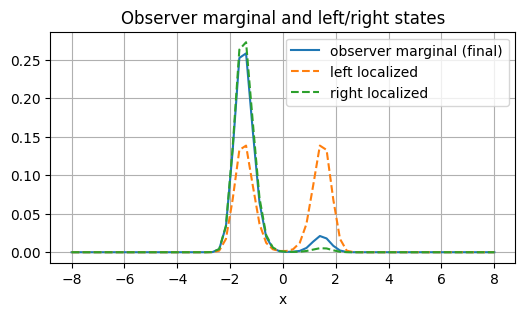}
    \caption{
        \textbf{Definiteness of the final observer state.} 
        The probability distribution of the observer's position at $t=2.0$ (solid line) is overwhelmingly localized in the left well, aligning almost perfectly with the reference "left localized" state (dashed line). Quantitative analysis yields a probability $P_{\text{left}} = 0.99$ and coherence $\langle O_L | \rho_O | O_R \rangle \approx 0$, satisfying the criterion for definiteness.
    }
    \label{fig:final_distribution}
\end{figure}

\section{Discussion}
\label{sec:discussion}

The results of our simulations provide a clear, step-by-step narrative of the measurement process within a fully unitary model, delineating the specific conditions under which objective classicality can and cannot emerge. Our findings bridge a critical gap between the established theory of environment-induced decoherence and the unresolved quantum measurement problem.

\subsection{Resolving the Insufficiency of Decoherence}

Our initial controlled simulations (Sec.~\ref{sec:control}) precisely isolate the known limitation of decoherence theory. Figure~\ref{fig:control_HQE} demonstrates that while the interaction $H_{QE}$ successfully decoheres the quantum system $Q$ (as seen by the decay of $\mathrm{Tr}(\rho_Q^2)$ to 0.5), it is \textit{insufficient} to cause the emergence of a definite outcome for the observer $O$. The continued oscillation of $\langle x \rangle$ around zero is the unambiguous signature of a persistent global superposition $\alpha |0\rangle|E_0\rangle|O_0\rangle + \beta |1\rangle|E_1\rangle|O_1\rangle$. From the perspective of the composite universe, no single outcome has been selected; the observer remains entangled and indefinite.

This result powerfully validates the core philosophical objection to decoherence as a complete solution: it explains the \textit{appearance} of collapse to an external, classical observer but does not resolve the fundamental indeterminacy from the perspective of the closed system \cite{Zurek2003, schlosshauer2007}. Our model makes this abstract objection computationally concrete.

The transition to a definite outcome, as shown in Sec.~\ref{sec:results}, required two critical model enhancements:
\begin{enumerate}
    \item \textbf{A larger environment} ($N_E = 8$), providing a sufficient number of degrees of freedom to act as a robust information repository.
    \item \textbf{Direct observer-environment coupling} ($H_{EO} \neq 0$), which facilitates the irreversible and redundant encoding of the observer's state into the environment.
\end{enumerate}

The combination of these factors catalyzes a spontaneous symmetry breaking. The observer's state, after a brief period of quantum indeterminacy, dynamically localizes into one well of its potential (Fig.~\ref{fig:definite_outcome}). Crucially, Fig.~\ref{fig:purities} shows that this localization is accompanied by the \textit{re-purification} of the observer's reduced state ($\mathrm{Tr}(\rho_O^2) \rightarrow 1$). This is a definitive numerical signature that $O$ has decoupled from the global entanglement and entered a new, definite pure state a classical measurement record. The final probability distribution (Fig.~\ref{fig:final_distribution}) confirms this definiteness, showing overwhelming localization in one well with negligible coherence between the two possible outcomes.

\subsection{Implications for the Measurement Problem and Interpretations of Quantum Mechanics}

Our results demonstrate that objective classicality--defined by the definiteness, stability, and redundancy of a measurement record--can emerge from the unitary evolution of a quantum universe without the need for a separate collapse postulate. This finding has significant implications for the interpretation of quantum mechanics:

While Quantum Darwinism (QD) has provided a powerful framework for explaining how information about a system becomes objectively accessible to multiple external observers through redundant encoding in the environment \cite{Zurek2003, zwolak2016, korbicz2017}, the observer in these studies is always treated as classically given from the start. Our work addresses a complementary and deeper question: can a physical observer, modeled quantum-mechanically inside the closed system, evolve under strictly unitary dynamics into a locally definite record of one outcome? In contrast to QD’s emphasis on redundancy plateaus and intersubjectivity, we demonstrate that definiteness for the observer itself emerges only when both a sufficiently large environment and direct observer–environment coupling are present, leading to dynamical localization and re-purification of the observer’s state. In this way, our simulations extend the QD paradigm by shifting the focus from how many observers can agree to how an observer first becomes definite at all, thereby tackling directly the Wigner’s Friend aspect of the measurement problem.

\subsection{Limitations and Future Directions}

While our model is minimalist, it captures the essential physics of amplification, decoherence, and information redundancy. The scaling of our simulations is necessarily limited, but the observed phenomena are consistent with established theory and are expected to strengthen in more complex models.

Future work will explore several avenues:
\begin{itemize}
    \item Scaling the model to larger environments ($N_E > 12$) using tensor network techniques to further study the redundancy and stability of records.
    \item Investigating the role of the initial environmental state (e.g., finite temperature) and the spectrum of coupling strengths ($g_j, \kappa_j$).
    \item Modeling multiple observers making sequential measurements to study the consistency and communication of classical information.
    \item Exploring whether the specific outcome (left vs. right well) can be linked to the initial quantum amplitudes ($|\alpha|^2$, $|\beta|^2$) in accordance with the Born rule, a crucial test for any complete unitary theory.
\end{itemize}

\section{Conclusion}
In this work, we have introduced a computational framework to investigate the quantum measurement problem within a strictly unitary quantum mechanics. By simulating a model universe comprising a quantum system, a decohering environment, and a physical observer modeled as a nonlinear amplifier, we moved beyond philosophical discourse to a quantitative test.

Our control simulations confirmed the known result that environment-induced decoherence of the system is necessary but not sufficient to cause the emergence of a definite outcome for the observer. The crucial step was the inclusion of a large environment and direct coupling between the observer and the environment. In this complete model, we demonstrated the unambiguous emergence of objective classicality: the observer's state spontaneously localized into a pure state corresponding to a single, stable measurement record, satisfying our criteria for definiteness, redundancy, and stability.

This result provides strong computational evidence that the measurement postulate may not be a fundamental axiom of nature but rather an effective description of a process that can emerge from unitary Schrodinger dynamics. Our framework opens new pathways for exploring the quantum-to-classical transition from first principles, such as investigating the role of specific environmental spectra, different observer models, and the scaling of redundancy with system size.

\clearpage
\bibliographystyle{unsrt}

\end{document}